\documentclass[letterpaper,review]{elsarticle}

\usepackage{latexsym}
\usepackage{amsmath,amssymb,amsfonts,amscd}
\usepackage[mathscr]{eucal}

\usepackage{eufrak}
\usepackage{graphicx}
\usepackage{bm}
\usepackage{amsmath}
\usepackage{amssymb}
\usepackage{color}
\usepackage{subfigure}
\usepackage{rotating}

\newcommand{\abinitio}{\textit{ab initio} }
\newcommand{\twobytwo}{$2\times 2\times 2$ }
\newcommand{\sixteenby}{$16\times 16\times 16$ }
\newcommand{\pv}{\ensuremath{\vec{p}} }
\newcommand{\pdif}[2]{ \ensuremath{ \frac{\partial #1}{\partial #2} } }

\newcommand{\PdotE}{\ensuremath{\vec{p}\cdot\vec{E}}}

\journal{Computer Physics Communications}

\begin{document}

\begin{frontmatter}

\title{A Modified Histogram Method for Disordered Lattices}

\author{Dennis L. Jackson}

\author{David Roundy\corref{cor}}
\cortext[cor]{Corresponding author. 1-541-250-0137}
\ead{roundyd@physics.oregonstate.edu}
\address{Department of Physics, Oregon State University, Corvallis, OR
  97331, USA}

\begin{abstract}
  A new method is developed to extend the histogram method
  \cite{ferrenberg1988new,ferrenberg1989optimized} to lattices with
  any type of disorder.  The Monte Carlo
  single- and multiple-histogram methods were developed to get the most out
  of only a few simulations, but are restricted to simulations with
  identical lattice configurations.  The method introduced here expands on
  the histogram method to allow data from various disordered
  lattice configurations to be optimally combined in a single weighted average
  result.  This method is applied to a simple Ising-like model of the
  relaxor ferroelectric perovskite solid solution $BaTiO_3$ -
  $Bi(Zn_{1/2}Ti_{1/2})O_3$ (BT-BZT), in which disorder plays a
  pivotal role.
\end{abstract}


\end{frontmatter}


\section{Introduction} 
Studying phase transitions using Monte Carlo simulations can be
challenging.  Various analysis methods have been developed, especially for
simple systems such as the Ising\cite{ising1925beitrag, onsager1944crystal,
  jensen1983mean} , Potts\cite{potts1952some, berg2004dynamics,
  bazavov2008phase}, and Heisenberg models\cite{peczak1991high,
  chen1993static}.  These methods can be expanded to more complex systems,
but there are limitations.  One thing that most Ising-like models have in
common is a homogeneous lattice, where each cell has the same accessible
magnetic states.  A useful tool for analyzing Monte Carlo data for systems
like this is the histogram method developed by Ferrenberg and
Swendsen\cite{ferrenberg1988new,ferrenberg1989optimized}.  The
histogram method allows for only a few simulations to be performed, while
maintaining a high level of accuracy in the calculations.  As long as the
necessary accessible states are sufficiently sampled, the accuracy of the
calculations remains high.  For a traditional Ising model, the histogram method
uses a histogram that is populated from accessible energy and magnetization
states to calculate the probability that the system will end up in a
particular state.  Several studies using the histogram
methods\cite{peczak1991high, ferrenberg1995statistical,
  ferrenberg1991critical, chen1993static} have shown promise in
improving the temperature resolution without increased computational time.

A modified histogram method is introduced to analyze phase transitions
of an Ising-like model in which the system is disordered and contains
cells with different accessible states.  One drawback of the histogram
method is that it is constrained to combining data from simulations of
homogeneous lattices of the same size.  The new method allows
simulations of differing lattice size to be combined in a similar
fashion, as well as allowing for the combination of inhomogeneous
disordered lattices.

The modified method is tested on a system---a solid solution relaxor
ferroelectric---containing cells with electric dipoles, instead of the
traditional magnetic dipoles.  Therefore, the measurables of the
system are energy and polarization, $E$ and $\vec{p}$.  The method
applies for computation of any thermal average of a function of $E$
and $\pv$ can be calculated.  See Section~\ref{sec:Application} for
more details on the solid solution used as a test case.  In this
context, the histograms will be populated with energies and
polarizations, instead of the usual energies and magnetizations.

\section{Single Histogram Method}
A conventional Monte Carlo simulation involves generating a set of $N$
samples that represent the canonical ensemble at a temperature $T$.
From this set of samples, the mean value of any function of the energy
and polarization may be found using the equation
\begin{equation}
  \langle f(E,\pv) \rangle = \frac{1}{N}\sum_{i=1}^N f(E_i,\pv_i) \;,
  \label{eq:simple_mean}
\end{equation}
where the subscript $i$ is the index of a particular data sample, and $N$
is the total number of samples.  Unfortunately, this method only gives
results at the temperature used to create the sampled data.  Thus, a
computationally expensive Monte Carlo simulation must be run for each
temperature of interest.  Using the single histogram method, the data from
a simulation at one temperature may be used to predict properties at
other similar temperatures\cite{ferrenberg1988new}.

The histogram method was created to extract as much information as
possible from a Monte Carlo simulation, while minimizing the
computational effort.  This technique relies on the fact that if the
probability of finding the system in each microstate at a given
temperature is precisely known, then in principle the probability of
finding the system in any of those microstates at any other
temperature can be predicted.  In practice, these probabilities can
only be approximated, and calculated temperatures must be relatively
close to simulated temperatures in order to achieve statistically
significant results.

The histogram method begins with constructing a histogram of microstates
$H(E,\pv)$, from a Monte Carlo simulation at a single temperature.  For the
simple Ising model, this histogram has a natural resolution in terms of
both energy and magnetization, with the resolution determined by the change
in each of these quantities due to a single cell flip.  In this paper,
consideration is given to disorder in the number of polarization states
available in each cell, complicating the choice of bin size.  These bin
size effects are discussed in more detail in Section~\ref{sec:binsize}.
The summations below are summations over the energy and polarization bins
of the histogram.

The single histogram method is used to calculate the probability of
finding the system in a particular microstate at a similar desired
temperature $T$, different from the temperature $T_0$ at which the
simulation was performed.  The probability is given by
\begin{equation}
  P_\beta(E,\pv) = \frac{1}{Z_\beta}H(E,\pv)\exp(-(\beta-\beta_0)(E-\PdotE) )\;,
\end{equation}
where $\beta_0 = 1/(k_BT_0)$, $\beta = 1/(k_BT)$, $\vec{E}$ is the
external applied electric field, and $Z_\beta$ is a the partition
function given by
\begin{equation}
  Z_\beta = \sum_{E,\,\pv}H(E,\pv)\exp(-(\beta-\beta_0)(E-\PdotE) )\;.
  \label{eq:hist_Z}
\end{equation}
The mean value of any quantity at the desired temperature is then found
using,
\begin{equation}
  \langle f(E,\pv) \rangle_\beta = \sum_{E,\,\pv}f(E,\pv)P_\beta(E,\pv)\;.
\label{eq:hist_mean}
\end{equation}
This technique is accurate if the desired temperature $T$ is near the
simulation temperature $T_0$, but becomes increasingly imprecise as
the temperature $T$ deviates from $T_0$, and the population
$P_\beta(E,\pv)$ becomes increasingly different from $H(E,\pv)$.
Details of this uncertainty are discussed in Section~\ref{sec:uncertainty}.

\section{Multiple Histogram Method}
When computing the value of a property over a large range of temperatures,
the single histogram method becomes sub-optimal.  Each calculation uses data
from only a single simulation temperature histogram.  When examining a
temperature precisely between two simulation temperatures, it would be
ideal to use data from both of those simulations.  The multiple-histogram
method addresses this issue, and allows for the use of all the Monte Carlo
simulation data for each computed data point.  Similar uncertainties are
achieved with similar computational effort by either simulating many
temperatures with fewer iterations at each temperature, or by running
simulations at fewer temperatures with more iterations in each simulations.
Thus the multiple-histogram method serves as an optimal interpolation
scheme for examining intermediate temperatures.  This allows for more
precise results, and leads to greatly reduced computation time.

The multiple histogram method as proposed by Ferrenberg and Swendsen
\cite{ferrenberg1989optimized}, uses the following equation for the
probability function at temperature $T$:
\begin{equation}
  \tilde{P}_\beta(E,\pv) = \frac{\sum_{n=1}^N H_n(E,\pv) g_n^{-1} \exp\left(-\beta
    E\right) } {\sum_{m=1}^N n_mg_m^{-1} \exp\left(-\beta_m E - f_m\right)}\;,
  \label{eq:P}
\end{equation}
where $\beta_m = 1/k_BT_m$, $N$ is the number of simulations,
$H_m(E,\pv)$ is the histogram of the $m^{th}$ simulation, $n_m$ is the
total number of samples included in the $m^{th}$ histogram, $g_m =
1+2\tau_m$ with $\tau_m$ being the correlation time, and $f_m$ is an
estimate of the dimensionless free energy of the $m^{th}$ simulation,
discussed below.  The correlation time serves as a weighting factor to
account for the differing correlation times of the samples for
different simulations, which affects the number of statistically
distinct configurations present.  The free energies serve as a
correction factor due to combining data from multiple temperatures,
and are calculated self consistently using
\begin{equation}
  \exp\left(f_m\right) = \tilde{P}_{\beta_m}(E,\pv)\;.
\end{equation}

\begin{figure}
  \centering 
  \subfigure[Dielectric constant with and without multiple-histogram
    method]{\includegraphics[width=\linewidth]
    {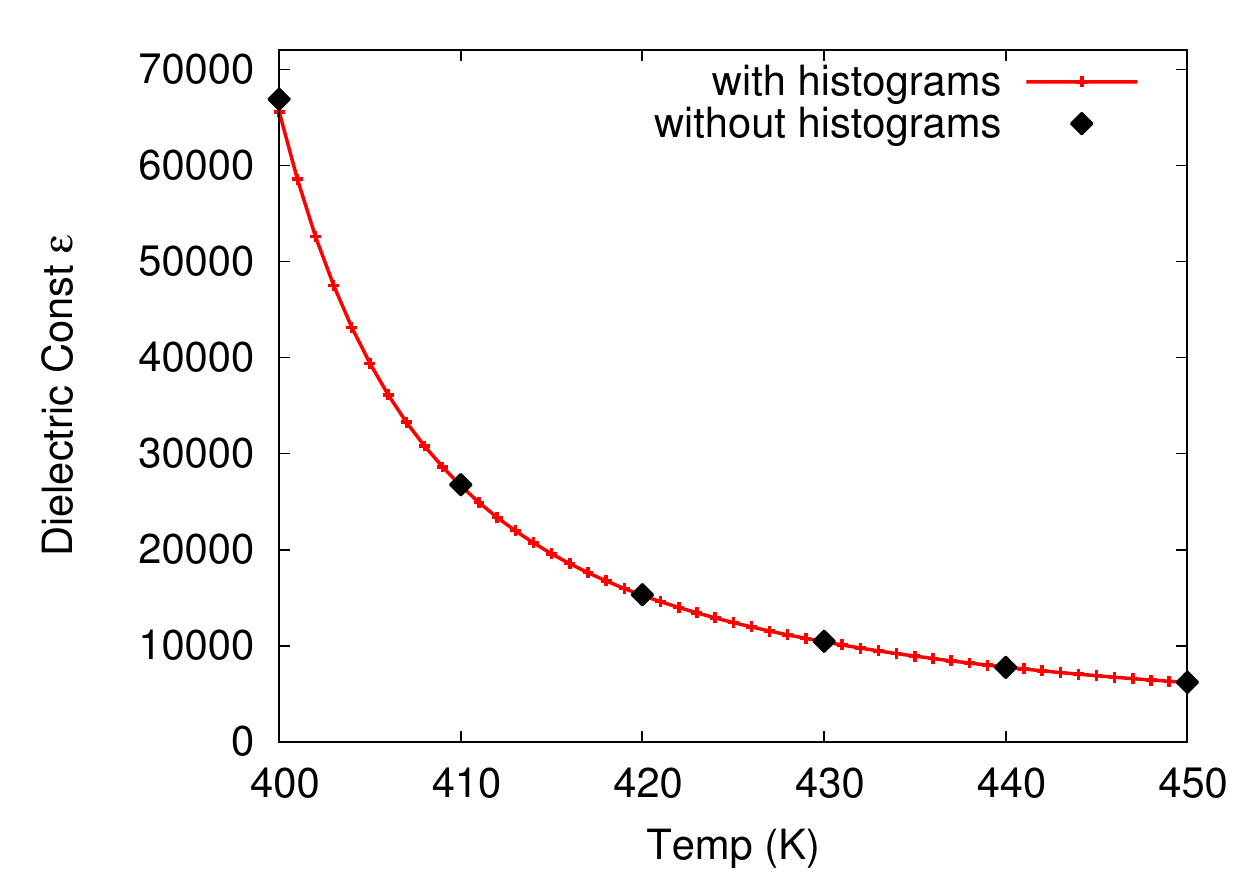}
    \label{fig:XvsT_with-without_hist_T400-10-450}}\\
  \subfigure[Specific heat with and without multiple-histogram
    method]{\includegraphics[width=\linewidth]
    {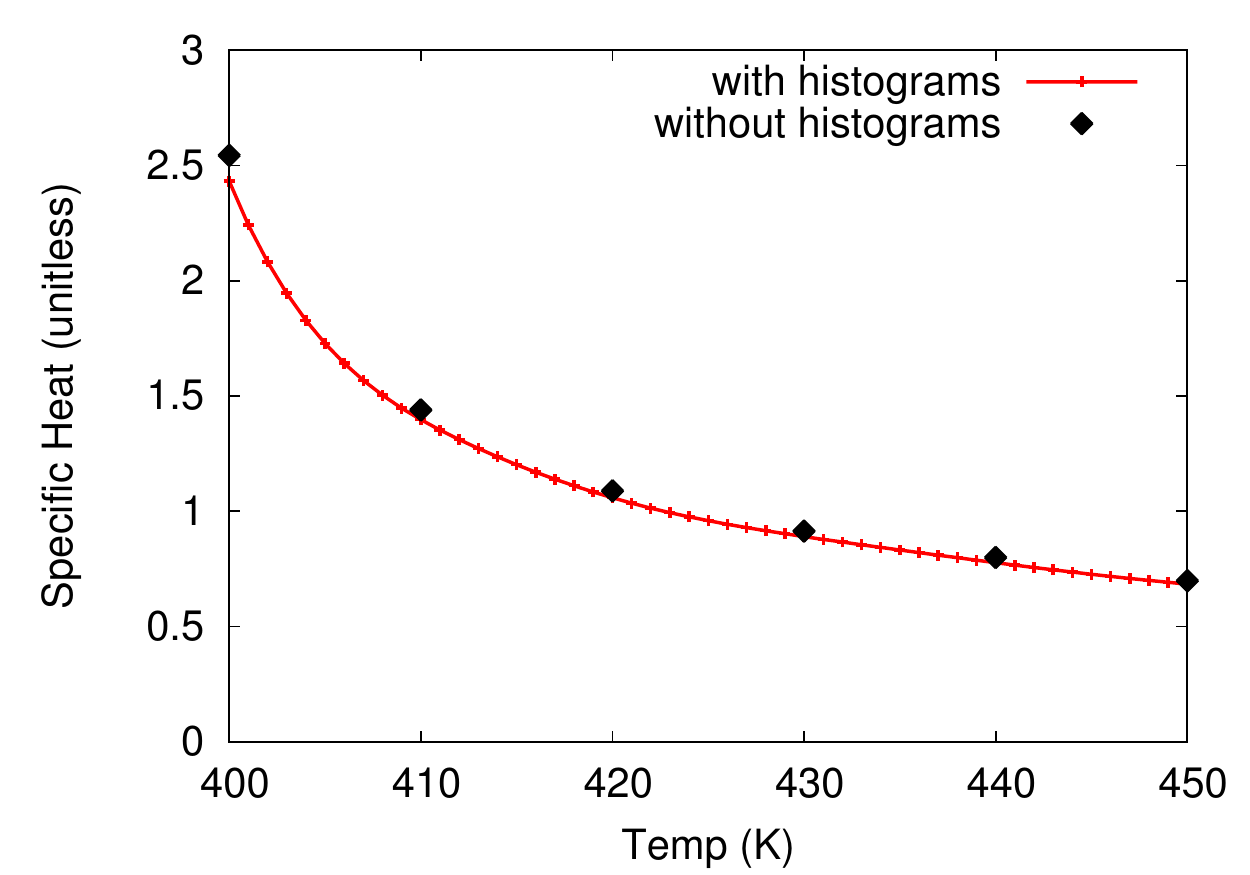}
    \label{fig:CvsT_with-without_hist_T400-10-450}}
  \caption[Dielectric constant and Specific Heat of a homogeneous
    \sixteenby lattice using raw data and the multiple histogram
    method.]{[Color online] Dielectric constant and Specific Heat of a homogeneous
    \sixteenby lattice using raw data and the multiple histogram method.
    The simulations used for the non-histogram method are also the inputs
    in the multiple histogram method.  The multiple histogram method is
    sampled every 1K, while the input simulations were performed every
    10K.}
  \label{fig:simple_data_vs_histograms}
\end{figure}

Note that Equation~\ref{eq:P} does not give the actual probability,
but a probability distribution function that is unnormalized.  To normalize
the distribution function and obtain an actual probability, it is just a
matter of dividing by another partition function:
\begin{equation}
  P_\beta(E,\pv) =
  \frac{\tilde{P}_\beta(E,\pv)}{\sum_{E,\pv}\tilde{P}_\beta(E,\pv)}\;.
  \label{eq:normalized_prob}
\end{equation}
Once the probabilities are determined, the mean value of a quantity is
once again calculated using Equation~\ref{eq:hist_mean}.  To illustrate the
interpolation benefits of the multiple-histogram method,
Figure~\ref{fig:simple_data_vs_histograms} compares the dielectric constant
and specific heat of a $16\times 16\times 16$ homogeneous lattice for
simulations at every 10 Kelvin and the multiple-histogram method using the
data from the same input simulations.  The specific heat and electric
susceptibility are essentially the variance of the energy and polarization
respectively:
\begin{equation}
  c = \frac{\beta^2}{N}\left(\langle E^2 \rangle - \langle E \rangle ^2\right)\;,
\end{equation}
where $N$ is the number of cells in the lattice, and
\begin{equation}
  \chi = \beta\left(\langle P^2 \rangle - \langle P \rangle ^2\right)\;.
  \label{eq:chi}
\end{equation}
Note that $\chi$ is a tensor value, and only a single independent direction may
be calculated at once.  The dielectric constant may also be calculated
using
\begin{equation}
  \epsilon = 1+4\pi\chi\;.
\end{equation}

\section{Histogram Bin Size}
\label{sec:binsize}

The bin size that determines the allowable microstates for energy
and polarization is not as easily defined for inhomogeneous lattices as it
is for the simple Ising model.  For the inhomogeneous model, instead of
calculating every possible microstate of energy and polarization, a
standard bin size is used because different lattice configurations may have
different sets of available energy and polarization microstates.  This also
allows for data from multiple simulations to be combined in a
straight-forward manner.  The bin size used here is determined by
calculating the energy and polarization differences for a single flip of
the cells with the smallest polarization.  The histogram is then filled
with the simulation data according to the smallest bin sizes.

Larger bin sizes are less computationally expensive than small bins, since
the mean value of any function of $E$ and $\pv$ is a sum over the available
microstates.  More bins means more microstates, and more microstates amount
to more computational time.  However, there is a limit to the maximum size
a bin may be before the technique loses accuracy.  The smallest bin size
that should be used is that which corresponds to the difference in energy
or polarization due to flipping a single cell from one state to another.
The dielectric constant is the variance of the polarization, and thus
depends only on the bin size of the polarizations.  Similarly, the specific
heat only depends on the bin size of the energy.  A simple test for the
application used here reveals that the maximum accurate bin size for both
energy and polarization occurs at about four times the minimum bin size.

\section{Combining Histograms of Different Configurations}
\label{sec:combined_histograms}
The multiple-histogram method addresses the possibility of combining
together simulations at several temperatures in order to reduce the
statistical errors associated with the number of iterations simulated.
When studying a disordered material such as a solid solution, there is
another sort of statistical error, which arises from the lattice size.
This could be addressed by simulating a very large lattice, but the
statistical error only drops as the square root of the size of the lattice.
If the lattice size of a completed simulation is determined to be too
small, it is discouraging to start again with a larger lattice, losing the
computational effort on the original lattice.  The new approach discussed
here combines simulations from distinct lattices that differ in disorder as
well as lattice size, so as to provide improved statistical averaging over
all disordered configurations.  While the conventional multiple-histogram
method combines simulations at different temperatures, the new modified
histogram method combines simulations in different lattice configurations
and temperatures, to provide better statistical averaging of disordered
materials.

Consider the case of two systems with the same lattice size that represent
the same system, with different disordered lattice configurations.  The two
systems may be combined into one large system with twice the volume and
number of cells of each original system.  To get accurate results for this
larger system, the probabilities are calculated for the new larger system
and are used to obtain the averaged values from
Equation~\ref{eq:hist_mean}.  Instead of running the full Monte Carlo
simulation of the larger cell, the probability of finding each microstate
of the large cell is approximated using the probability distributions of
the smaller subsystems.

In general, the probability of finding multiple configurations each
with a particular energy is the product of the individual probabilities
\begin{equation}
    P_{tot}(E_1,E_2,\cdots,E_n) =
    P_1(E_1) P_2(E_2) \cdots P_N(E_n)\;,
\label{eq:P_tot_1}
\end{equation}
where $P_i(E_{j_i})$ is the probability of finding system $i$ with
energy $E_i$, under the assumption that the systems may be considered
non-interacting and statistically independent.  For brevity, the
discussion here is only focused on the energy of the system, but the
method is trivially extended to include polarization or
magnetization.

To describe a system that combines multiple systems together into a
larger volume, the resulting probability must be a function of the total
energy, rather than the energy of each subsystem as in
Equation~\ref{eq:P_tot_1}.  This is accomplished by integrating over all
energy microstates with an appropriate delta function constraining the
total energy
\begin{align}
    P_{tot}(E_{tot}) &= \int dE_1\int dE_2 \cdots \int dE_n
    P_1(E_{j_1})P_2(E_{j_2})\cdots P_N(E_{j_n})\delta(E-(E_1+E_2+\cdots+E_n))\;.
\end{align}
Considering only discrete states and
including the polarization dimension as well, the combined probability
becomes
\begin{equation}
  \begin{split}
    P_{\beta,tot}\left(E,\pv\right) = \sum_{E_1,\pv_1}\sum_{E_2,\pv_2}
    \cdots \sum_{E_n,\pv_n}
    P_{\beta,1}\left(E_1,\pv_1\right)P_{\beta,2}\left(E_2,\pv_2\right)
    \cdots P_{\beta,N}\left(E_N,\pv_N\right) \\
    \delta_{E , (E_1+E_2+\cdots+E_N)}
    \delta_{\pv , (\pv_1+\pv_2+\cdots+\pv_N)}\;,
  \end{split}
  \label{eq:combined_prob}
\end{equation}
where the subscript $\beta$ indicates the dependence of probability on
temperature.
  
One challenge that arises when calculating the total probability this
way is that it is computationally expensive.  If each sum is order
$\mathcal{O}(M)$, where $M$ is the number of energy and polarization
microstates for an individual lattice configuration, and the total
probability is a combination of $N$ lattice configurations, the resulting
probability calculation is order $\mathcal{O}(M^N)$.  This can
be a serious performance issue when combining more than just a few lattice
configurations.  Convolutions are used to speed up the computational
process.

A convolution of two functions $f(t)$ and $g(t)$ is defined as 
\begin{equation}
  (f\circ g) (t) \equiv \int_{-\infty}^\infty f(\tau)g(t-\tau)d\tau\;.
\end{equation}
To quickly calculate convolutions, the convolution theorem is utilized:
\begin{equation}
  \mathcal{F}[f\circ g] = \mathcal{F}[f] \cdot \mathcal{F}[g]\;,
\end{equation}
where $\mathcal{F}$ represents a Fourier transform.  Recognizing the
similarity between the new total probability, given in
Equation~\ref{eq:combined_prob}, and a series of convolutions allows for
the use of the advantageous discrete Fourier transforms to calculate the
final unnormalized probability:
\begin{equation}
  \begin{split}
    \tilde{P}_{\beta,tot}\left(E,\pv\right) &=
    \mathcal{F}^{\;-1}\Big[\mathcal{F}[P_{\beta,1}\left(E_1,\pv_1\right)]
      \times \\ & \mathcal{F}[P_{\beta,2}\left(E_2,\pv_2\right)] \times
      \cdots \times\mathcal{F}[P_{\beta,N}\left(E_N,\pv_N\right)]
      \Big]\;.
  \end{split}
\end{equation}
The final normalized probability is found using
Equation~\ref{eq:normalized_prob}.

In order to use the Fourier transform technique, each individual
probability must have the same number of microstates as the total
probability.  As was previously stated, the number of microstates of the
final probability is of order $\mathcal{O}(M^N)$.  This number can be
dramatically reduced by making the bins of each individual histogram, and
subsequently each polarization, conform to the same microstate grid.  If
each histogram has the same bin size and is centered on the same values,
the number of microstates of the final polarization can be reduced to
\begin{align}
  M_{E_{total}} &= \Big[M_{E_1} + M_{E_2} +\cdots + M_{E_N} - (N-1)\Big]\;,\\
  M_{\pv_{total}} &= \Big[M_{\pv_1} + M_{\pv_2} + \cdots + M_{\pv_N} -
    (N-1)\Big]\;,\\
  M &= M_{E_{total}}\times M_{\pv_{total}}\;,
\end{align}
where $M_{E_i}$, $M_{\pv_i}$ are the number of energy and polarization
microstates of the $i^{th}$ lattice configuration respectively, and $N$ is
the number of lattice configurations being combined.  To calculate the
total probability of Equation~\ref{eq:combined_prob}, there are $N$
convolutions required and $3N$ FFTs per convolution.  Each FFT is of order
$\mathcal{O}(M \log M)$, so the total probability calculation is of order
$\mathcal{O}(N\times M\log M)$.  When combining many simulations, or even a
few simulations with many bins, this can be computationally intensive.


\section{Uncertainties}\label{sec:uncertainty}
The level of accuracy in any computed quantity is always of
great importance, and the multiple histogram method is no exception.  In
general, Monte Carlo methods have the advantage of begin able to produce
rigorous statistical error bars.  The statistical uncertainty of the multiple
histogram probability is given by
\begin{equation}
  \delta P_{\beta}\left(E,\pv\right)=
  \left(\sum_{n=1}^N \frac{H_n\left(E,\pv\right)}{g_n}\right)^{-1/2}
  P_{\beta}\left(E,\pv\right)\;,
  \label{eq:dP}
\end{equation}
where the sum is over all $N$ simulations, $H_n(E,\pv)$ is the histogram of a
single simulation, and $g_n = 1+2\tau_n$ is the correlation time factor
\cite{ferrenberg1989optimized}.

\begin{figure}
  \centering 
  \includegraphics[width=\linewidth]{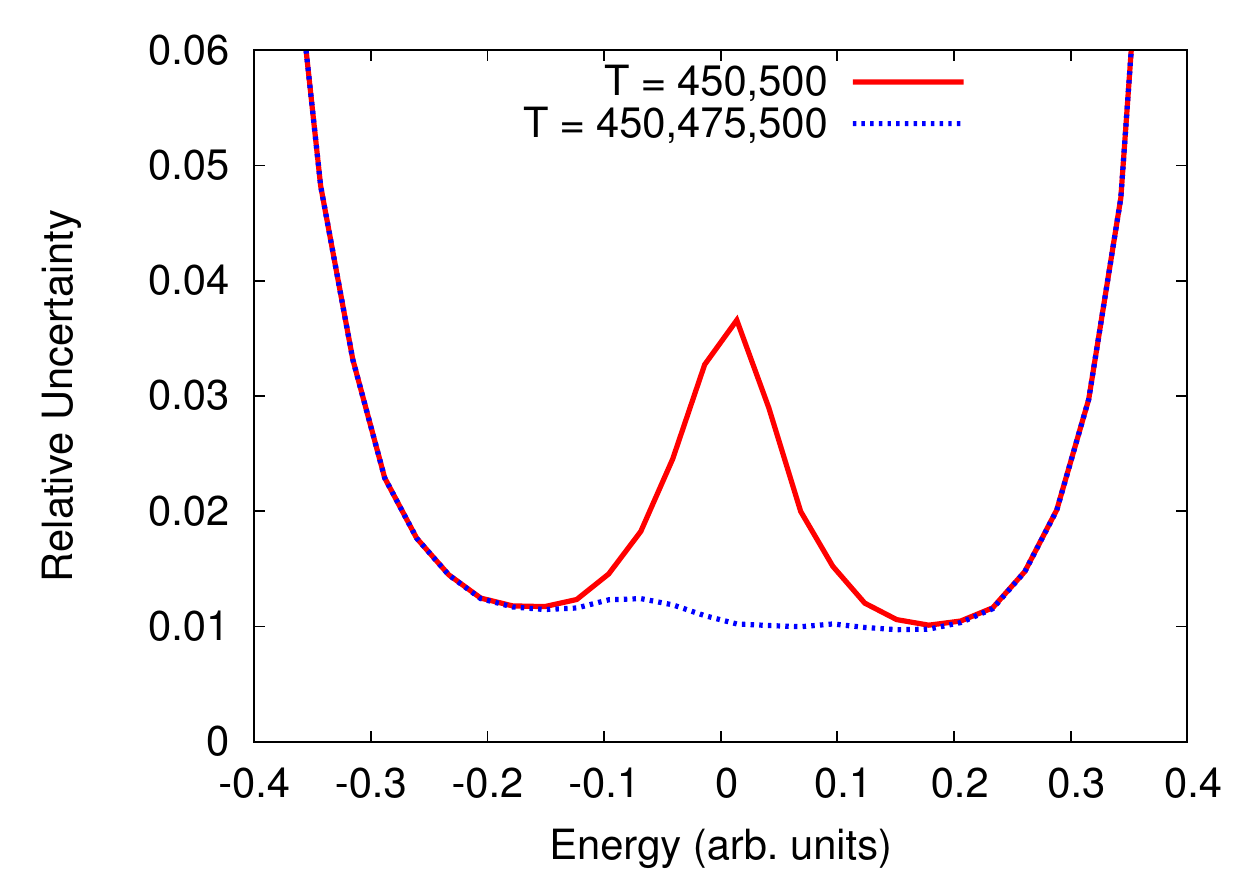}
  \caption[The relative uncertainty for two different multiple
    histograms.]{[Color online] The relative uncertainty for two different
    multiple histograms. One with simulations at T=450,500K, and the other
    with simulations at T=450,475,500K.  The relative uncertainty with only
    two simulations has a region of higher uncertainty.  Adding a third
    simulation to the histogram data lowers the uncertainty in the center
    region so the relative uncertainty of the probability for the entire
    valid region is less than 2\%.  }
  \label{fig:relative_uncertainty}
\end{figure}

The form of Equation~\ref{eq:dP} is similar to that of any counting
experiment, with a relative uncertainty $(\delta P/P)$ proportional to
$1/\sqrt{N}$.  Just as more counts reduce the uncertainty in any normal
counting experiment, more counts per bin of each histogram reduce the
statistical error in the probabilities.  If the relative uncertainty
is high, more Monte Carlo data must be obtained to fill in the gaps for
those microstates.  The relative uncertainty should be low for the entire
range of occupied microstates in order to produce reliable results at a
given temperature.  Figure~\ref{fig:relative_uncertainty} shows the
relative uncertainty of Equation~\ref{eq:dP} using both two and three
simulation temperatures.  The relative uncertainty with only two
simulations has a region of high uncertainty, while adding a third
simulation reduces the uncertainty for a larger temperature range.  The
uncertainty at the end points diverges as there are no occupied microstates
far from the simulated temperatures.

\begin{figure}
  \centering 
  \includegraphics[width=\linewidth]{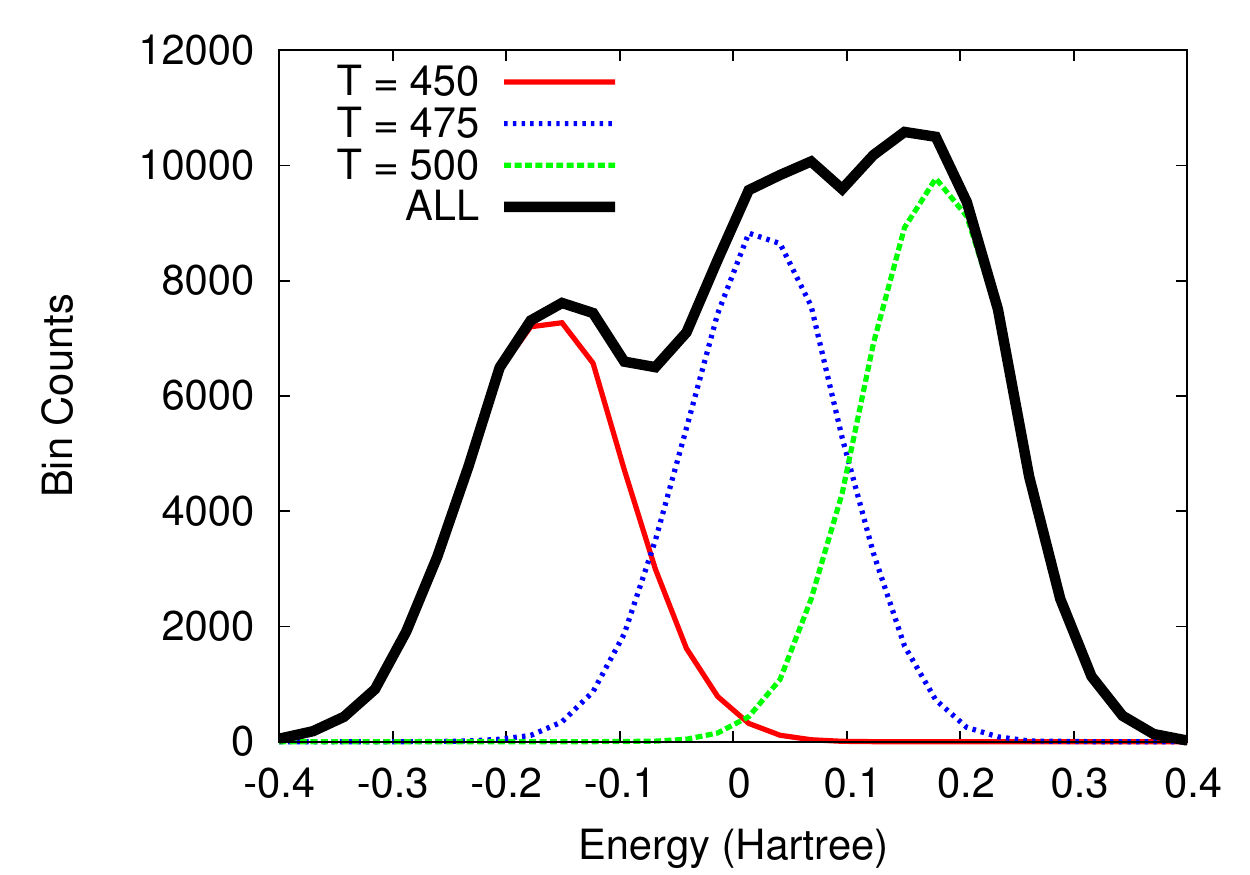}
  \caption[Single and multiple energy histograms]{[Color online] Single and
    multiple energy histograms.  Each of the three histograms have
    considerable overlap and combine to form one continuous multiple
    histogram.  Sufficient overlap of histograms reduces the finite sample
    error.}
  \label{fig:simple_histograms}
\end{figure}

Another source of error in histogram re-weighting is due to using a
finite number of samples in the histogram.  For an excellent
discussion of errors associated with the histogram method, see Newman
and Palmer\cite{newman1999error}, who discuss this error in detail.
As long as the histograms of every simulation temperature have
sufficient overlap, the finite sample error will be small for the
entire temperature range.  Figure~\ref{fig:simple_histograms} shows
the overlap of three histograms at different simulation temperatures.
If the amount of overlap is sufficient, the finite sample error will
be small because the number of counts in each bin of the histogram is
large.  At the end points of the histogram the finite sample error
becomes large because the histogram is not filled in that region.
This becomes extremely important if the histogram contains empty bins.
According to Equation~\ref{eq:dP}, the relative statistical
uncertainty is infinite for microstates with empty bins.  To avoid
this problem, only data that has a reasonably high number of samples
is used.  Thus, the statistical error will dominate over finite
sampling errors and remain finite.  This may require more simulations
or longer runs.

Equation~\ref{eq:dP} gives the statistical error for the usual
multiple-histogram method of Ferrenberg and
Swendsen\cite{ferrenberg1989optimized}.  When combining multiple
lattice configurations, the uncertainty equation must be modified in a
way that is consistent with the combined probability of
Equation~\ref{eq:combined_prob}.  To make this clear, a shorthand for
the probability in Equation~\ref{eq:combined_prob} is introduced as
\begin{equation}
  P_{tot} = \sum_1\sum_2 \cdots \sum_N P_1 P_2 \cdots P_N \;,
  \label{eq:short_combined_prob}
\end{equation}
where each summation is over the microstates of the probabilities of the
same subscript, and the delta functions are simply implied.  The
uncertainty in the total probability is then given by
\begin{align}
  \begin{split}
    \delta P_{tot}^2 = \sum_1\left( \pdif{P}{P_1} \delta P_1 \right)^2 &+
    \sum_2\left( \pdif{P}{P_2} \delta P_2 \right)^2 + \cdots \\
    &+ \sum_N\left( \pdif{P}{P_N} \delta P_N \right)^2\;,
    \label{eq:dP_multi}
  \end{split}
\end{align}
where the uncertainty for each lattice configuration $\delta P_i$ is
calculated using Equation~\ref{eq:dP}.  This method is valid if each
probability is independent, that is, the errors in probabilities are
uncorrelated.

The derivative of Equation~\ref{eq:short_combined_prob} with respect to a
single lattice configuration probability is
\begin{equation}
  \pdif{P}{P_i} = \sum_1 \cdots \sum_{i-1} \sum_{i+1} \cdots \sum_N P_1
  \cdots P_{i-1} P_{i+1} \cdots P_N \;.
\end{equation}

The error in the total probability is then given by
\begin{equation}
  \begin{split}
    \delta P_{tot}^2 = \sum_1\Bigg\lbrack\left( \delta P_1 \right)^2\left( 
    \sum_2\sum_3\cdots\sum_N P_2 P_3 \cdots P_N \right)^2\Bigg\rbrack + \\ 
    \sum_2\Bigg\lbrack\left( \delta P_2 \right)^2\left( 
    \sum_1\sum_3\cdots\sum_N P_1 P_3 \cdots P_N \right)^2\Bigg\rbrack + \cdots\\ 
    + \sum_N\Bigg\lbrack\left( \delta P_N \right)^2\left( 
    \sum_1\sum_2\cdots\sum_{N-1} P_1 P_2 \cdots P_{N-1} \right)^2\Bigg\rbrack \;.
  \end{split}
  \label{eq:dP_multi_2}
\end{equation}
While this method is accurate considering the assumption that each
probability is independent, it is computationally challenging,
even when FFTs are used to do the convolutions.  Since convolutions are
essentially three discrete Fourier transforms, each convolution scales as
that of an FFT, which is of order $\mathcal{O}{(M\log M)}$, with $M$ being
the number of microstates and histogram bins.  The uncertainty of the
combined probability requires $N^2$ convolutions, thus the computational
time is order $\mathcal{O}{(N^2 \times M\log M)}$.

Once the total uncertainty $\delta P$ is found, the uncertainties in the
polarization magnitude, specific heat, and dielectric constant may be
determined.  Here, the results are summarized for the uncertainty of each
of the thermodynamic averages that are computed.  First, consider the
polarization magnitude.  The uncertainty in the mean magnitude of
polarization is given by
\begin{equation}
  \delta\langle \,|\,\pv\, |\,\rangle^2 = \sum_{E,\pv}\left(
  \pdif{\langle\, |\,\pv \,|\,\rangle}{P_\beta(E,\pv)} \delta
  P_\beta(E,\pv)\right)^2\;.
  \label{eq:dpmag}
\end{equation}
Taking the derivative of Equation~\ref{eq:hist_mean} results in
\begin{equation}
  \delta\langle \,|\,\pv\, |\, \rangle^2 = \sum_{E,\pv}\left(\,|\,\pv\, |\,
  - \langle \,|\,\pv\, |\,\rangle\right)^2\left(\frac{\delta P}{Z}\right)^2\;.
  \label{eq:dpmag_2}
\end{equation}
The uncertainties of the susceptibility can be determined in the same
manner as Equation~\ref{eq:dpmag},
\begin{equation}
  \delta\chi^2 =
  \sum_{E,\pv}\left(\frac{\partial\chi\left(\pv\right)}{\partial
    P_\beta\left(E,\pv\right)}\delta P_\beta\left(E,\pv\right)\right)^2 .
  \label{eq:dchi}
\end{equation}
By carrying out the derivative of Equation~\ref{eq:chi}, and using the
average values in the form of Equation~\ref{eq:hist_mean}, the resulting
uncertainty in the susceptibility is given by
\begin{equation}
  \begin{split}
    \delta\chi^2 = \frac{\beta\,^2}{Z_\beta^2}\sum_{E,\pv}
    \Big\lbrack\big\lbrack\pv\;^2 - \langle\pv\;^2\rangle +
    2\langle\pv\rangle\,^2 - 2\langle\pv\rangle\,\pv\big\rbrack^2 \times\\
    \left(\delta P_\beta\left(E,\pv\right)\right)^2\Big\rbrack \;,
  \end{split}
\end{equation}
and similarly for the specific heat,
\begin{equation}
  \begin{split}
    \delta c^2 = \frac{\beta\,^4}{M^2 Z_\beta^2}\sum_{E,\pv} \Big\lbrack\big\lbrack
    E\,^2 - \langle E\,^2\rangle + 2\langle E\rangle\,^2 - 2\langle E\rangle
    E\big\rbrack\,^2 \times \\ \left(\delta
    P_\beta\left(E,\pv\right)\right)^2\Big\rbrack\;,
  \end{split}
\end{equation}
where $M$ is the number of cells on the lattice.


\section{Application}
\label{sec:Application}
The material that inspired this method is the solid solution
$x\text{BaTiO}_3 +
(1-x)\text{Bi}(\text{Zn}_{1/2}\text{Ti}_{1/2})\text{O}_3$ (BT-BZT).
For certain compositions, BT-BZT is a relaxor ferroelectric
perovskite, which exhibits both long- and short-range disorder.  At
these compositions, BT-BZT features the so-called diffuse phase
transition.  For other compositions of BT-BZT behaves as a
ferroelectric such as pure $BaTiO_3$, which maintains an ordinary
ferroelectric phase transition.

The relaxor behavior is modeled using a combination of techniques
including \abinitio methods and an Ising-like model.  The \abinitio
calculations are accurate at small length scales, but lack the ability to
describe the necessary long-range interactions.  The Ising-like model is
able describe long-range effects, but cannot predict short-range
interactions.  By combining these methods into a single model, insight is
gained about the larger picture by including both long- and short-range
effects.

The \abinitio calculations are restricted to \twobytwo supercells of BT-BZT
for symmetry reasons.  This allows for compositions of $x = 0, 0.25, 0.5,
0.75, 1$, in the 40 atom supercell.  Calculations are performed to determine
the energies and polarizations of each unique configuration using the
Quantum Espresso \cite{QE-2009} package, Density Functional Theory (DFT)
and the Modern Theory of Polarization (MTP) \cite{king1993theory,
  vanderbilt1993electric, resta2007theory}.  The DFT calculations were run
using a PBE-GGA exchange-correlation functional, ultrasoft
pseudopotentials, a cutoff energy of 80 Rydberg, and a \twobytwo
Monkhorst-Pack $k$-point grid within the supercell.  For all but $x=1$
there are multiple supercell configurations, each with multiple discrete
polarization states.  This disorder prompted the development of the new
modified histogram method.  For supercells with several polarization, each
such state is treated within the Monte Carlo simulations.

After the energies and all polarization states of each \twobytwo
configuration are determined for all compositions, this information is
used to construct an Ising-like model.  The classic Ising model for
ferromagnetic systems has two states per lattice position: spin up and spin
down\cite{ising1925beitrag, onsager1944crystal, jensen1983mean}.  The
Heisenberg model improves this model by introducing a continuum of states,
with the spin pointing in any direction.  In this model, a lattice of
\twobytwo supercells is used, which is populated with the atomic
configurations discussed above.  Each state has a set of possible
polarizations determined by the \emph{ab initio} calculations.  The number
of each type of \twobytwo supercells are determined using Boltzmann
statistics at a formation temperature, and are distributed on the lattice
stochastically.  This differs from both Ising and Heisenberg models in that
the polarizations are not simply up and down, but neither are they allowed
to point in any direction.  This adaptation is more closely related to the
Potts model \cite{potts1952some, berg2004dynamics, bazavov2008phase} in
that there are several discrete directions of the polarizations, but unlike
the Potts model the polarizations of the disordered lattice are not uniform
in nature and each cell on the lattice may have different accessible
polarization states.

The interaction Hamiltonian of the Heisenberg model used here is given by
\begin{equation}
  H = -\frac{J}{2} \sum_{i}^{cells}\sum_{j}^{NN} \vec{p}_i\cdot\vec{p}_j 
  - \sum_{i}^{cells}\vec{p}_i\cdot\vec{E}\; ,
  \label{eq:hamiltonian}
\end{equation}
where $\vec{p}_i$ is the polarization of a single cell, $\vec{E}$ is
the external electric field and $J$ is a coupling constant.  The
coupling constant is tuned to match the experimental Curie temperature
of pure barium titanate using the fourth order Binder cumulant method
\cite{Binder}.  This Hamiltonian does not take into account all of the
long-range forces, but it is a good first approximation.
\begin{figure}
  \subfigure[]{
    \includegraphics[width=\linewidth]{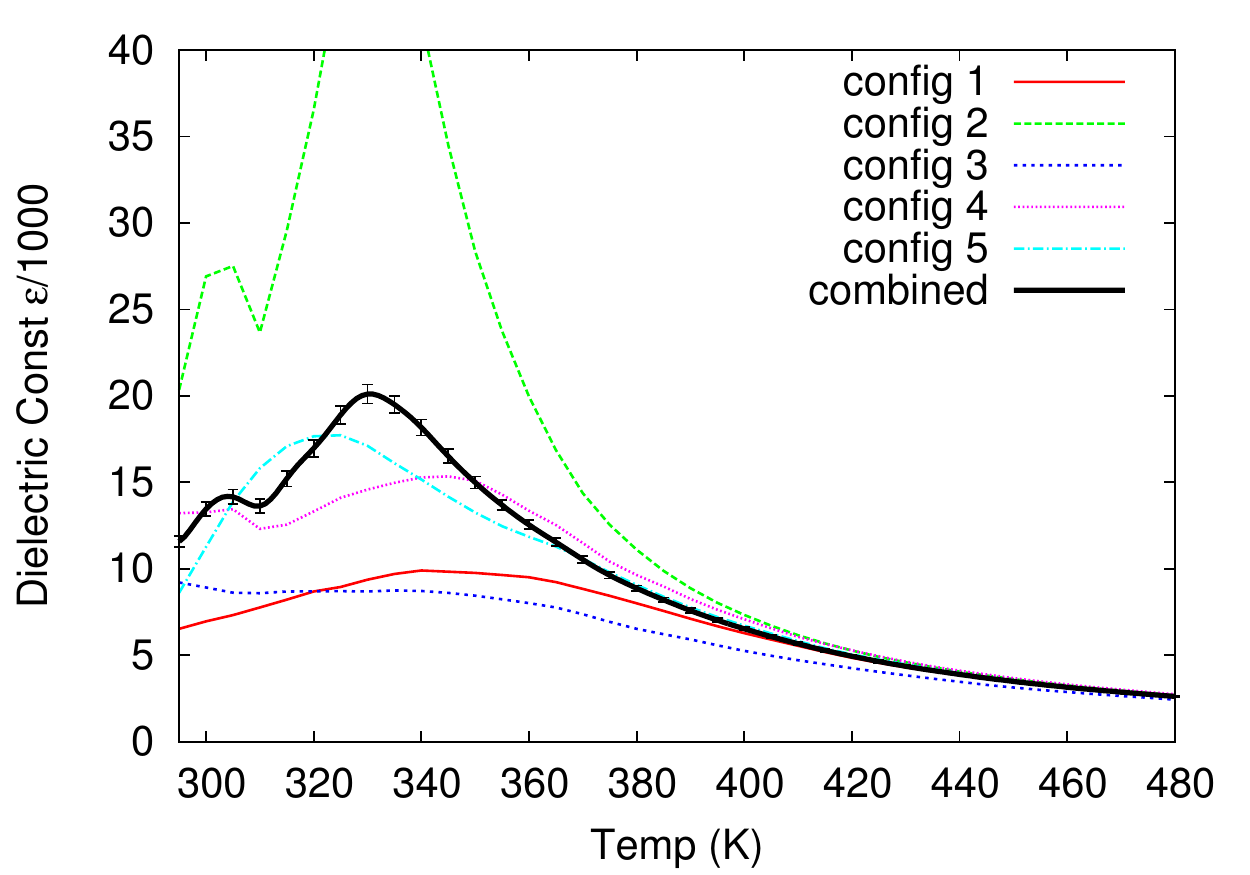}
  }
  \subfigure[]{
    \includegraphics[width=\linewidth]{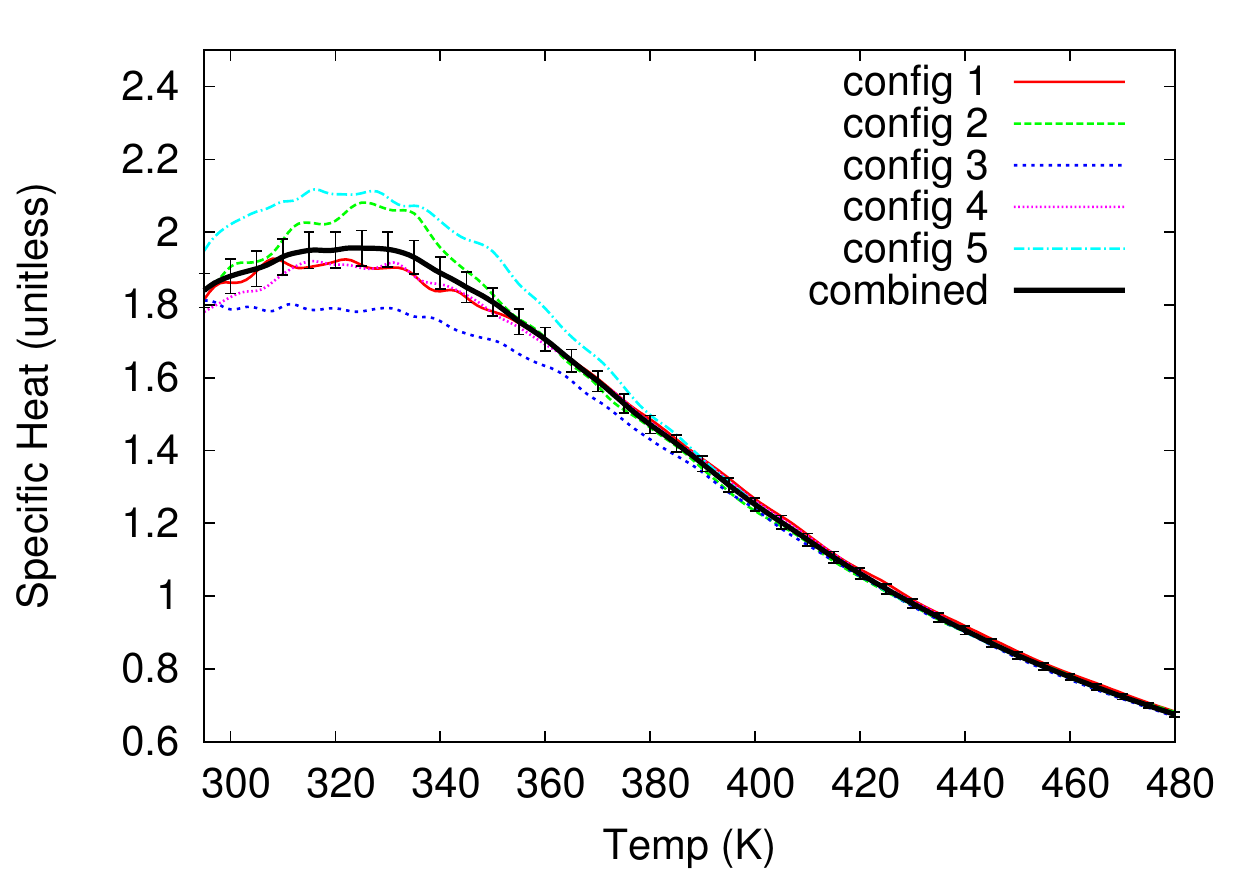}
  }
  \caption{[Color online] The (a) dielectric constant and (b) specific
    heat for five different lattice configurations and the combined
    result with error bars for a composition of $x=0.95$.}
  \label{fig:XCvsT_N16_ext12345_r1_with_errorbars}
\end{figure}

Figure~\ref{fig:XCvsT_N16_ext12345_r1_with_errorbars} shows the dielectric
constant and specific heat for five different lattice configurations of
BT-BZT, each with a composition of $x = 0.95$, as well as the convolved
combination of each configuration.  Pure BTO ($x=1$) has a distinct phase
transition near 380 K.  As expected for a relaxor, the data in
Figure~\ref{fig:XCvsT_N16_ext12345_r1_with_errorbars} correctly shows that
a sharp phase transition does not occur, and the material exhibits a
diffuse phase transition.  There is a large variation between individual
configurations because the accessible energy and polarization microstates
are different for each lattice configuration.

The combined result behaves very much like a weighted average of each
individual result for both the specific heat and the dielectric
constant.  This is unsurprising, given that it corresponds to system
composed of precisely these configurations.  At high temperatures,
where the results from each lattice configuration are very similar,
the convolved results are quite smooth.  At lower temperatures, there
is quite a difference in the results for each lattice configuration,
as the system freezes into different states according to the precise
disordered state.  The convolved results begin to smooth out compared
to the results for an individual simulation, but there is a higher
degree of uncertainty, and adding additional lattice configurations
would reduce the uncertainty.

\section{Conclusion and Summary}
A modification of the multiple-histogram method for analysis of Monte Carlo
data has been introduced.  This new modification allows for easier and more
efficient study of systems with a disordered lattice.  The method was
applied to a model of the relaxor ferroelectric perovskite solid solution
$BaTiO_3$ - $Bi(Zn_{1/2}Ti_{1/2})O_3$ (BT-BZT).  A total of 
255 simulations were combined using the new modified histogram method to
compute the specific heat and dielectric constant of this disordered
material with improved accuracy and precision.  


\bibliographystyle{elsarticle-num}
\bibliography{thesis}

\begin{thebibliography}{10}
\expandafter\ifx\csname url\endcsname\relax
  \def\url#1{\texttt{#1}}\fi
\expandafter\ifx\csname urlprefix\endcsname\relax\def\urlprefix{URL }\fi
\expandafter\ifx\csname href\endcsname\relax
  \def\href#1#2{#2} \def\path#1{#1}\fi

\bibitem{ferrenberg1988new}
A.~Ferrenberg, R.~Swendsen, {New Monte Carlo technique for studying phase
  transitions}, Physical review letters 61~(23) (1988) 2635.

\bibitem{ferrenberg1989optimized}
A.~Ferrenberg, R.~Swendsen, {Optimized monte carlo data analysis}, Physical
  Review Letters 63~(12) (1989) 1195--1198.

\bibitem{ising1925beitrag}
E.~Ising, {Beitrag zur theorie des ferromagnetismus}, Zeitschrift f{\"u}r
  Physik A Hadrons and Nuclei 31~(1) (1925) 253--258.

\bibitem{onsager1944crystal}
L.~Onsager, {Crystal statistics. I. A two-dimensional model with an
  order-disorder transition}, Physical Review 65~(3-4) (1944) 117.

\bibitem{jensen1983mean}
M.~Jensen, P.~Bak, {Mean-field theory of the three-dimensional anisotropic
  Ising model as a four-dimensional mapping}, Physical Review B 27~(11) (1983)
  6853.

\bibitem{potts1952some}
R.~Potts, {Some generalized order-disorder transformations}, in: Mathematical
  Proceedings of the Cambridge Philosophical Society, Vol.~48, Cambridge Univ
  Press, 1952, pp. 106--109.

\bibitem{berg2004dynamics}
B.~Berg, H.~Meyer-Ortmanns, A.~Velytsky, {Dynamics of phase transitions: The 3D
  3-state Potts model}, Physical Review D 70~(5) (2004) 054505.

\bibitem{bazavov2008phase}
A.~Bazavov, B.~Berg, S.~Dubey, {Phase transition properties of 3D Potts
  models}, Nuclear Physics B 802~(3) (2008) 421--434.

\bibitem{peczak1991high}
P.~Peczak, A.~M. Ferrenberg, D.~Landau, High-accuracy monte carlo study of the
  three-dimensional classical heisenberg ferromagnet, Physical Review B 43~(7)
  (1991) 6087.

\bibitem{chen1993static}
K.~Chen, A.~Ferrenberg, D.~Landau, {Static critical behavior of
  three-dimensional classical Heisenberg models: A high-resolution Monte Carlo
  study}, Physical Review B 48~(5) (1993) 3249.

\bibitem{ferrenberg1995statistical}
A.~Ferrenberg, D.~Landau, R.~Swendsen, {Statistical errors in histogram
  reweighting}, Physical Review E 51~(5) (1995) 5092.

\bibitem{ferrenberg1991critical}
A.~Ferrenberg, D.~Landau, {Critical behavior of the three-dimensional Ising
  model: A high-resolution Monte Carlo study}, Physical Review B 44~(10) (1991)
  5081.

\bibitem{newman1999error}
M.~Newman, R.~Palmer, {Error estimation in the histogram Monte Carlo method},
  Journal of Statistical Physics 97~(5) (1999) 1011--1026.

\bibitem{QE-2009}
P.~G. et. al., \href{http://www.quantum-espresso.org}{{QUANTUM ESPRESSO: a
  modular and open-source software project for quantum simulations of
  materials}}, Journal of Physics: Condensed Matter 21~(39) (2009) 395502
  (19pp).
\newline\urlprefix\url{http://www.quantum-espresso.org}

\bibitem{king1993theory}
R.~King-Smith, D.~Vanderbilt, {Theory of polarization of crystalline solids},
  Physical Review B 47~(3) (1993) 1651.

\bibitem{vanderbilt1993electric}
D.~Vanderbilt, R.~King-Smith, {Electric polarization as a bulk quantity and its
  relation to surface charge}, Physical Review B 48~(7) (1993) 4442.

\bibitem{resta2007theory}
R.~Resta, D.~Vanderbilt, {Theory of polarization: a modern approach}, Physics
  of Ferroelectrics (2007) 31--68.

\bibitem{Binder}
K.~Binder, {Critical properties from Monte Carlo coarse graining and
  renormalization}, Physical Review Letters 47~(9) (1981) 693--696.

\end{thebibliography}

\end{document}